\documentclass[final,3p,times]{elsarticle}

\usepackage{amssymb}
\usepackage{amsmath}

\usepackage{multirow}

\usepackage{hyperref}

\usepackage{svg}
\usepackage{subfigure}
\usepackage{comment}

\usepackage{tikz}
\usetikzlibrary{3d,arrows.meta,decorations.markings}
\usetikzlibrary{backgrounds,automata}
\usepackage{tikz-3dplot}

\usepackage{mathptmx}
\usepackage{amsmath, bm}

\begin{document}

\begin{frontmatter}

\author[1]{Mohd. Meraj Khan} 

\affiliation[1]{organization={Department of Applied Mechanics and Biomedical Engineering, Indian Institute of Technology Madras},
            city={Chennai},
            country={India}}

\author[2]{Sumesh P. Thampi} 

\affiliation[2]{organization={Department of Chemical Engineering, Indian Institute of Technology Madras},
            city={Chennai},
            country={India}}

\author[1]{Anubhab Roy\corref{cor1}}
\ead{anubhab@iitm.ac.in}
\ead[URL]{https://home.iitm.ac.in/anubhab}
\cortext[cor1]{}

\title{Lattice Boltzmann method for electromagnetic wave scattering}

\begin{abstract}

In this work, the lattice Boltzmann method (LBM) is assessed as a time-domain numerical approach for electromagnetic wave scattering. Owing to its explicit formulation and suitability for parallel computation on structured grids, LBM provides an alternative framework for solving Maxwell’s equations. The formulation is first validated using canonical benchmarks, including reflection and refraction at a planar dielectric interface and two-dimensional scattering from infinitely long circular cylinders, where the computed angular scattering intensities are compared with analytical Lorenz--Mie solutions. Additional comparisons are performed for circular cylinders with varying dielectric constants to examine performance across different material contrasts. The framework is then extended to three-dimensional scattering from dielectric spheres, representing the most computationally demanding case considered in this work, and the resulting angular scattering intensities are compared with exact Lorenz--Mie solutions. To further examine performance for non-circular geometries, scattering from an  infinitely long hexagonal dielectric cylinder is investigated and benchmarked against results obtained using the Discretized-Mie Formalism. Across all cases, the LBM predictions show close agreement with analytical and semi-analytical reference solutions over a range of size-to-wavelength ratios.

\end{abstract}



\end{frontmatter}



\section{Introduction}
\label{sec:intro}

Electromagnetic scattering is a fundamental wave phenomenon in which incident radiation interacts with obstacles and redistributes energy into different directions \cite{kerker1969scattering,liou2016light}. Accurate prediction of scattered fields is essential in applications ranging from atmospheric optics and remote sensing to radar technology, astrophysics, biomedical imaging, and nanophotonics \cite{kahnert2016numerical}. For canonical geometries such as spheres and infinitely long circular cylinders, exact solutions are available through Lorenz–Mie theory and related formulations \cite{hulst1981light,stratton2007electromagnetic}. In contrast, most practical scatterers exhibit irregular shapes, heterogeneous compositions, or sharp geometrical features, for which closed-form solutions are not available and numerical solvers of Maxwell’s equations must be employed \cite{mishchenko2014electromagnetic}.

Among the numerical approaches developed for such problems, the discrete dipole approximation (DDA) and the finite-difference time-domain (FDTD) methods are widely used for arbitrarily shaped particles \cite{kahnert2016numerical}. The DDA is formulated in the frequency domain, representing the scatterer as an array of interacting dipoles, whereas FDTD operates in the time domain by directly discretizing Maxwell’s curl equations in space and time.

The lattice Boltzmann method (LBM), originally introduced as a mesoscopic numerical method for fluid dynamics, has more recently been extended to computational electromagnetics \cite{mendoza2010three, liu2014lattice, hauser2017stable}. Rooted in kinetic theory \cite{kruger2017lattice, mohamad2011lattice}, LBM computes electromagnetic fields as moments of mesoscopic distribution functions that evolve through discrete streaming and collision processes \cite{hauser2017stable}. Unlike FDTD, which directly discretizes Maxwell's curl equations in space and time, LBM is derived from a kinetic description whose moments recover the macroscopic fields — making it a fundamentally distinct numerical framework rather than a refinement of conventional field-based discretizations, despite both methods being explicit time-domain schemes operating on structured grids. A quantitative comparison by Hauser and Verhey \cite{hauser2019comparison} showed that, for equal spatial resolution, LBM entails higher memory usage and computational cost due to the additional distribution functions. However, when evaluated at comparable error levels, LBM requires fewer lattice points, partially offsetting its higher per-node overhead. These findings motivate further investigation of the LBM framework for electromagnetic scattering problems, which is the focus of the present work.

In our previous work \cite{khan2024electromagnetic}, LBM was applied to two-dimensional (2D) electromagnetic scattering problems involving dielectric and perfect electrically conducting (PEC) circular cylinders of infinite length (axially invariant geometries). That study focused primarily on near-field distributions and radiation force calculations and did not include systematic validation of far-field scattering observables.

To the best of our knowledge, apart from Ref.~\cite{khan2024electromagnetic}, there has been no comprehensive benchmarking study devoted specifically to electromagnetic scattering using LBM. In particular, validation against analytical Lorenz–Mie solutions for canonical three-dimensional (3D) geometries, as well as against semi-analytical formulations for sharp-edged scatterers, has not been systematically examined.

The present work addresses these gaps by providing a structured validation of LBM for electromagnetic scattering in one-, two-, and three-dimensional configurations. We first consider the fundamental one-dimensional (1D) problem of reflection and refraction at a planar dielectric and magnetic interface and compare LBM predictions with analytical solutions. We then examine 2D scattering from PEC and dielectric circular cylinders, corresponding to infinitely long geometries invariant along the axial direction, and benchmark the results against Lorenz–Mie theory. This is followed by a regular hexagonal dielectric cylinder in a 2D configuration, validated against the Discretized-Mie Formalism (DMF). Finally, we investigate 3D scattering from dielectric spheres and compare the results with exact Lorenz–Mie solutions. A schematic overview of these test cases is provided in Fig.~\ref{fig:schematic}.

\begin{figure}[htb]
\centering
\begin{tikzpicture}[scale=1.1, >=stealth]

\node at (0,1.4) {(a) Planar interface (1D)};
\fill[gray!20] (0,-1.2) rectangle (1.8,0.2); 
\draw[thick] (0,-1.2) -- (0,0.2); 
\foreach \y in {-1.0,-0.7,-0.4,-0.1} {
  \draw[->,blue,thick] (-1.5,\y) -- (0,\y);
}
\foreach \y in {-1.0,-0.7,-0.4,-0.1} {
  \draw[->,red,thick] (0,\y+0.15) -- (-1.2,\y+0.15);
}
\foreach \y in {-1.0,-0.7,-0.4,-0.1} {
  \draw[->,green!60!black,thick] (0,\y) -- (1.5,\y);
}

\node at (4,1.4) {(b) Circular cylinder (2D)};
\draw[thick, fill=gray!20] (4,-0.5) circle (0.5);
\foreach \y in {-1.0,-0.7,-0.4,-0.1} {
  \draw[->,blue,thick] (2.5,\y) -- (3.5,\y);
}
\foreach \angle/\len in {0/1.2, 45/0.9, 90/0.8, 135/0.7, 180/0.5, 215/0.7, 270/0.8, 315/0.9} {
  \draw[->,red,thick] (4,-0.5) ++(\angle:0.55) -- ++(\angle:\len);
}
\foreach \y in {-0.6,-0.4} {
  \draw[->,green!60!black,thick] (3.7,\y) -- (4.3,\y);
}

\usetikzlibrary{shapes.geometric}

\node at (8,1.4) {(c) Hexagonal cylinder (2D)};

\node[draw, thick, fill=gray!20, regular polygon, regular polygon sides=6, shape border rotate=90, minimum size=1.2cm] (hex) at (8,-0.5) {};

\foreach \y in {-1.0,-0.7,-0.4,-0.1} {
  \draw[->,blue,thick] (6.5,\y) -- (7.3,\y);
}

\foreach \angle/\len in {0/1.0, 45/0.7, 90/0.7, 135/0.7, 180/0.5, 215/0.7, 270/0.7, 315/0.7} {
  \draw[->,red,thick] ([shift=(\angle:0.6cm)]hex.center) -- ++(\angle:\len);
}
\foreach \y in {-0.1,0.1} { 
  \draw[->,green!60!black,thick] ([xshift=-0.3cm,yshift=\y cm]hex.center) -- ++(0:0.6);
}

\node at (12,1.4) {(d) Sphere (3D)};
\shade[ball color=gray!50] (12,-0.5) circle (0.6);
\foreach \y in {-1.0,-0.7,-0.4,-0.1} {
  \draw[->,blue,thick] (10.5,\y) -- (11.5,\y);
}
\foreach \angle/\len in {0/1.0,60/1.3,120/0.9,180/0.6,240/0.8,300/1.2} {
  \draw[->,red,thick] (12,-0.5) ++(\angle:0.65) -- ++(\angle:\len);
}
\foreach \y in {-0.6,-0.4} {
  \draw[->,green!60!black,thick] (11.7,\y) -- (12.3,\y);
}

\end{tikzpicture}
\caption{Schematic of the scattering configurations considered: (a) planar dielectric interface, (b) circular cylinder (2D), (c) hexagonal cylinder (2D), and (d) sphere (3D). Blue, red, and green arrows denote the incident, scattered, and transmitted fields, respectively.}
\end{figure}\label{fig:schematic}

The remainder of the paper is organized as follows. Section~\ref{sec:method} outlines the LBM formulation for electromagnetic wave propagation and describes the computation of scattering observables such as the angular scattering intensity. Section~\ref{sec:results} presents validation results and numerical comparisons, and Section~\ref{sec:conclusion} summarizes the findings and discusses future extensions.

\section{Methodology}\label{sec:method}

\subsection{Lattice Boltzmann method framework}

The LBM, previously adapted for electromagnetic wave scattering in our earlier work \cite{khan2024electromagnetic}, is employed here to model the interaction between the incident field and the scatterer. This framework has also been recently applied to compute radiation forces and torques on particles with heterogeneous optical properties \cite{khan2025radiation}. Maxwell’s equations are solved on a D3Q7 lattice, where the distribution functions associated with the electric and magnetic fields are denoted by ${\bf e}_i({\bf r}, t)$ and ${\bf h}_i({\bf r}, t)$, respectively, with the subscript $i$ indexing the lattice velocity directions. These functions evolve according to the lattice Boltzmann update rules:
\begin{subequations} \label{eq:LBGK_equation}
\begin{gather}
    {\bf e}_{i}({\bf r} + {\bf c}_i\Delta t, t + \Delta t) =  2 {\bf e}_{i}^{eq}({\bf r} ,t) - {\bf e}_{i}({\bf r} ,t)  \\
    {\bf h}_{i}({\bf r} + {\bf c}_i\Delta t, t + \Delta t) = 2 {\bf h}_{i}^{eq}({\bf r} ,t) - {\bf h}_{i}({\bf r} ,t)
\end{gather}
\end{subequations}
where $\Delta t$ is the time step and ${\bf c}_i$ the lattice velocity.

The equilibrium distribution functions for the electric and magnetic fields, ${\bf e}_i^{eq}({\bf r}, t)$ and ${\bf h}_i^{eq}({\bf r}, t)$, are given as \cite{hauser2017stable}:
\begin{subequations} \label{eq:EDF}
\begin{gather}
    {\bf e}_i^{eq} ({\bf r} ,t) = 
    \begin{cases}
      \frac{1}{6}\Big(\pmb{ \mathcal{E}} ({\bf r} ,t) -   {\bf c}_i \times \pmb{ \mathcal{H}} ({\bf r} ,t) \Big) & \text{if $i \neq 0$}\\
      (\varepsilon_{r} - 1 ) \pmb{ \mathcal{E}} ({\bf r} ,t) & \text{if $i = 0$}
    \end{cases}  \\
    {\bf h}_i^{eq} ({\bf r} ,t) = 
    \begin{cases}
      \frac{1}{6}\Big(\pmb{ \mathcal{H}} ({\bf r} ,t) +  {\bf c}_i \times \pmb{ \mathcal{E}} ({\bf r} ,t) \Big) & \text{if $i \neq 0$}\\
      (\mu_{r} - 1 ) \pmb{ \mathcal{H}} ({\bf r} ,t) & \text{if $i = 0$}
    \end{cases}  
\end{gather}
\end{subequations}
where $\varepsilon_r$ and $\mu_r$ denote the dielectric constant and relative permeability, respectively. The vacuum permittivity $\varepsilon_0$ and permeability $\mu_0$ are set to unity. By applying the Chapman--Enskog expansion to Eq.~\eqref{eq:LBGK_equation} and substituting the equilibrium distribution functions from Eq.~\eqref{eq:EDF}, the Maxwell’s curl equations are recovered. The factor of 3 appearing in the recovered equations arises from the fact that, in lattice units, the speed of light in vacuum is $c = 1/3$ \cite{hauser2017stable}.

\begin{equation}
    \pmb{\nabla} \times \pmb{\mathcal{E}} = -3 \mu_r \frac{\partial \pmb{\mathcal{H}}}{\partial t}, \quad \pmb{\nabla} \times \pmb{\mathcal{H}} = 3 \varepsilon_r \frac{\partial \pmb{\mathcal{E}}}{\partial t}.
\end{equation}

The formulation above corresponds to non-absorbing media, which is the setting adopted throughout this work. Material absorption can in principle be incorporated by introducing an electrical conductivity term $\sigma\mathbf{E}$ in the magnetic curl equation \cite{hauser2017stable}, though the accuracy of this extension in multidimensional scattering configurations requires further investigation and is deferred to future work.

The macroscopic fields, $\pmb{\mathcal{E}}({\bf r}, t)$ and $\pmb{\mathcal{H}}({\bf r}, t)$, are reconstructed from the zeroth-order moments of ${\bf e}_i({\bf r}, t)$ and ${\bf h}_i({\bf r}, t)$ \cite{hauser2017stable}. The scattered fields are then obtained by subtracting the incident fields from the total fields. To suppress spurious reflections at the domain boundaries, open boundary
conditions are applied following the procedure described in Ref.~\cite{khan2024electromagnetic}
and summarized in Sec.~\ref{sec:OBC}. The LBM scheme employed here is known to be second-order accurate in both space and time, as demonstrated by Hauser and Verhey \cite{hauser2017stable, hauser2019comparison} through systematic error-scaling studies. This accuracy order is comparable to that of FDTD.

The above formulation is implemented in a hybrid numerical framework, designed to balance computational efficiency with coding flexibility. The core LBM solver is written in C to exploit OpenMP-based parallelization, while Python is used as the driver language for pre- and post-processing. The C routines are accessed within Python using the \texttt{ctypes} interface, which combines the efficiency of compiled C kernels with the ease of Python for coding, data handling, and visualization. All simulations were executed on multi-core CPUs, with the number of parallel computational threads indicated in the tables. Since simulations were run on different machines and thread counts were not systematically optimized (with higher thread counts sometimes chosen to reduce wall-clock time), the reported computation times should be regarded as indicative of scaling behavior rather than absolute benchmarks. In addition, the computational domain size varies with the size-to-wavelength ratio $a/\lambda$, with larger domains used for smaller values of $a/\lambda$. Since the LBM is a time-domain method, larger domains increase both the number of lattice points and the time required for waves to propagate across the domain, leading to longer simulation times even for small size-to-wavelength ratios.

While the LBM framework provides the time evolution of the electromagnetic fields within the computational domain, quantitative assessment of scattering requires translating these fields into measurable quantities. Because the simulations are performed in a finite domain, appropriate open boundary conditions must first be implemented to suppress artificial reflections from the boundaries (Sec.~\ref{sec:OBC}). Once the near-field solution is obtained, the scattering characteristics are quantified using far-field observables (Sec.~\ref{sec:RCS}). Since these observables are defined in the far-field limit, they are extracted from the simulated near-field data using a near-to-far-field (NTFF) transformation, described in Sec.~\ref{sec:NTFF}.


\subsection{Open boundary condition}\label{sec:OBC}

Since LBM simulations are performed in a finite computational domain, appropriate open boundary conditions are required to prevent artificial reflections from contaminating the scattered field. In the present work, non-reflecting boundary conditions are implemented following the procedure described in Ref.~\cite{khan2024electromagnetic}, which approximates outgoing-wave behavior at the domain boundaries.

At the outer boundary, the incoming distribution functions are approximated using a first-order extrapolation from the first interior node. Specifically, for a boundary node located at position $L$, the distribution function is taken as
\begin{equation}
f_i^{L} = f_i^{L-\Delta x},
\end{equation}
where $f_i$ represents a generic scalar distribution function corresponding to either the electric-field distributions ${\bf e}_{i}$ or the magnetic-field distributions ${\bf h}_{i}$. Here, $L$ denotes the boundary node and $L-\Delta x$ represents the first interior node adjacent to the boundary. This condition effectively assumes that the outgoing wave continues to propagate outward without reflection.

To assess the effectiveness of the open boundary treatment, we consider a validation test consisting of a 2D PEC circular cylinder placed in a square computational domain of side length $L = 4a$, where $a$ denotes the cylinder radius. The size-to-wavelength ratio is fixed at $a/\lambda = 1$. A sinusoidal pulse is introduced, and the total electromagnetic energy within the computational domain is monitored as a function of time for different spatial resolutions ($\lambda/\Delta x = 20$, 30, and 40).

\begin{figure*}[!htb]
\centering
\subfigure[Real part of the total electric field.]{\includegraphics[width=0.48\textwidth]{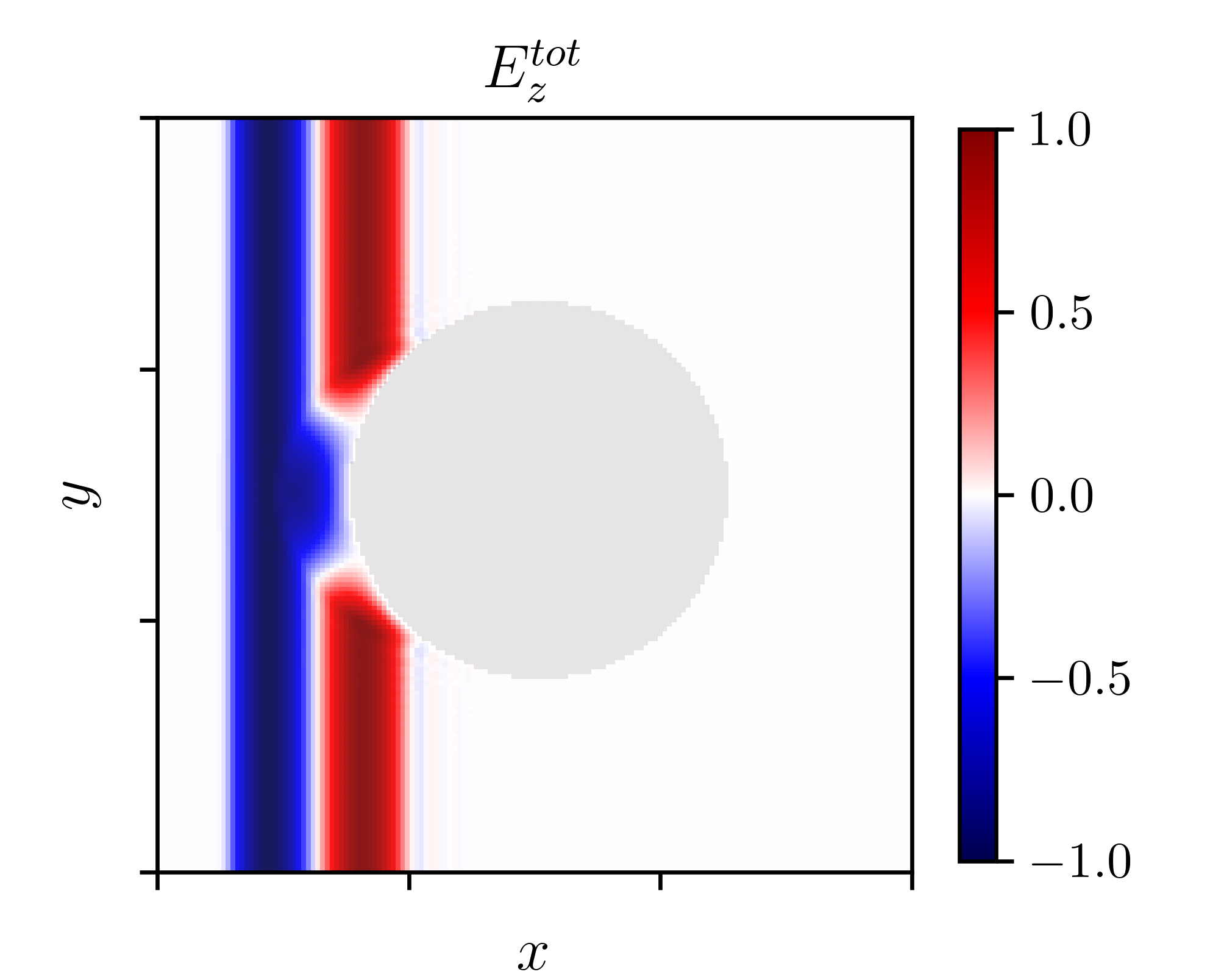}}
\subfigure[Normalized total electromagnetic energy versus time.]{\includegraphics[width=0.48\textwidth]{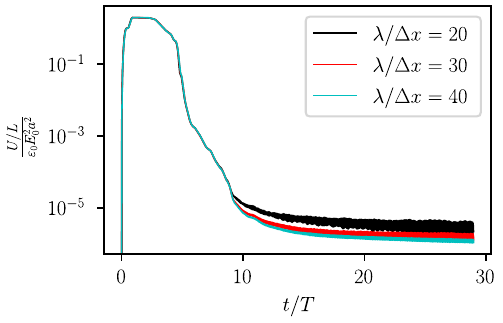}}
\caption{Validation of the open boundary condition for a 2D PEC circular cylinder at $a/\lambda = 1$ ($L = 4a$). (a) Real part of the total electric field. (b) Normalized total electromagnetic energy versus time for $\lambda/\Delta x = 20, 30,$ and $40$.}
\label{fig:open_BC}
\end{figure*}

Figure~\ref{fig:open_BC}(a) presents a snapshot of the total electric field as the pulse interacts with the 2D PEC circular cylinder. The temporal evolution of the normalized total energy is shown in Fig.~\ref{fig:open_BC}(b). During the interaction of the incident and scattered fields with the cylinder, the energy reaches a peak value close to 2 (in normalized units). As the pulse leaves the computational domain, the energy first decreases to approximately $10^{-3}$, indicating the departure of the main wave packet. Subsequently, the residual energy continues to decay and stabilizes at levels on the order of $10^{-5}$ or lower. A slight reduction of the residual energy is observed with increasing spatial resolution.

These results indicate that reflections from the domain boundaries remain small compared to the peak energy and decrease under grid refinement, demonstrating that the implemented open boundary conditions provide satisfactory radiation behavior for the configurations considered in this study.

\subsection{Far-field scattering intensity}\label{sec:RCS}

Having introduced the LBM formulation, we now describe the far-field observables used to quantify electromagnetic scattering. While LBM provides the time evolution of the electric and magnetic fields within the computational domain, scattering characteristics are most conveniently analyzed in terms of far-field quantities.

In the far-field region ($r \gg a$), where $r$ is the observation distance from the scatterer center and $a$ its characteristic size, the scattered field has the asymptotic form $e^{-ikr}/r$ in 3D and $e^{-ikr}/\sqrt{r}$ in 2D. This asymptotic behavior allows the definition of a far-field scattering amplitude $S$, whose squared magnitude characterizes the angular scattering distribution.

Using the asymptotic scaling of the scattered field, the angular scattering intensity is computed from the numerically obtained fields as
\begin{equation}
|S(\theta,\phi)|^2
=
k^2 r^2
\frac{|\mathbf{E}^S|^2}{|\mathbf{E}^I|^2},
\qquad \text{(3D)}
\label{eq:S3D}
\end{equation}

\begin{equation}
|S(\theta)|^2
=
k r
\frac{|\mathbf{E}^S|^2}{|\mathbf{E}^I|^2},
\qquad \text{(2D)}
\label{eq:S2D}
\end{equation}
where $k = 2\pi/\lambda$ is the free-space wavenumber, $\mathbf{E}^I$ denotes the incident electric field amplitude in the vicinity of the scatterer, and $\mathbf{E}^S$ the scattered electric field evaluated at a distance $r$ in the observation direction defined by the unit vector $\hat{\mathbf{r}}(\theta,\phi)$ (or $\hat{\mathbf{r}}(\theta)$ in 2D). The prefactors $k^2 r^2$ (3D) and $k r$ (2D) compensate for the radial decay, yielding a quantity that depends only on the scattering direction and is independent of the observation distance. The incident field considered in this work is linearly polarized; therefore, a single scattering amplitude $S$ characterizes the far-field response for the chosen polarization state. As these observables are defined in the far field, they cannot be obtained directly from the finite computational domain, and a NTFF transformation is therefore employed to extract the far-field quantities.

\subsection{Near-to-far-field transformation}\label{sec:NTFF}

To extract far-field quantities from the finite computational domain, we employ a NTFF transformation based on the equivalence principle, following Schneider and Taflove \cite{schneider2010understanding,taflove2005computational}.

\begin{figure}[htb]
\centering  

\subfigure[Near-field data recorded on a fictitious boundary.]  
{  
    \begin{tikzpicture}[scale=1.7,>=stealth]
        \fill[gray!40] (0,0) circle (0.35);
        \node at (0,0) {scatterer};

        \draw[thick] (-1.8,-1.8) rectangle (1.8,1.8);
        \draw[dashed,thick] (-1.1,-1.1) rectangle (1.1,1.1);
        \node[above] at (0,1.1) {fictitious boundary};

        \node[left] at (-1.1,0.35) {$\mathbf{E}_1,\mathbf{H}_1$};
        \node[left] at (-0.3,0.35) {$\mathbf{E}_2,\mathbf{H}_2$};
    \end{tikzpicture}

}  
\subfigure[NTFF transformation using equivalent surface currents.]  
{  

    \begin{tikzpicture}[scale=1.7,>=stealth]

        \draw[thick] (-1.8,-1.8) rectangle (1.8,1.8);
        \draw[dashed,thick] (-1.1,-1.1) rectangle (1.1,1.1);
        \node[above] at (-1.0,1.1) {fictitious boundary};

        \node[left] at (-1.1,0.35) {$\mathbf{E}_1,\mathbf{H}_1$};

        \node[left] at (-0.5,0.35) {$\mathbf{0},\mathbf{0}$};

        \draw[->,thick] (1.1,-0.7) -- (1.4,-0.7) node[right]{$\hat{\bf{n}}'$};

        \draw[->,thick] (1.1,0.0) -- (1.1,0.3);
        \node[right] at (1.15,0.15) {$\mathbf{J} = \hat{\bf{n}}' \times \mathbf{H}_1$};

        \draw[->,thick] (1.1,-0.45) -- (1.1,-0.15);
        \node[right] at (1.15,-0.35) {$\mathbf{M} = -\hat{\bf{n}}' \times \mathbf{E}_1$};

        \draw[->,thick] (0,0) -- (1.1,0.5);
        \draw[->,thick] (0,0) -- (2.7,1.9);
        \draw[->,thick] (1.1,0.5) -- (2.7,1.9);

        \draw (2.7,1.9) circle (1.5pt);

        \draw[dash pattern=on 6pt off 3pt,thick] (0,0) -- (1.4,0);
        \draw[dash pattern=on 6pt off 3pt,thick] (0,0) -- (0,1.4);

        \node[above] at (0,1.4) {y-axis};
        \node[above] at (1.6,-0.2) {x-axis};

        \node[above] at (0.7,0.5) {$\bf{r}$};
        \node[above] at (1.0,0.17) {$\bf{r'}$};
        \node[above] at (2.1,0.9) {$\bf{r} - \bf{r'}$};
        \node[above] at (2.9,1.5) {far point};

        \draw[->] (0.4,0) arc [start angle=0, end angle=38, radius=0.4];

        \node[right] at (0.35,0.1) {$\theta$};
        \node[right] at (0.55,0.13) {$\theta'$};

        \draw[->] (0.6,0) arc [start angle=0, end angle=25, radius=0.6];

    \end{tikzpicture}

}

\caption{Schematic of the NTFF transformation. (a) Near-field data recorded on a fictitious boundary enclosing the scatterer. (b) Computation of far-field quantities from equivalent surface currents.}\label{fig:NTFF_schematic}
\end{figure}

The basic idea is illustrated schematically in Fig.~\ref{fig:NTFF_schematic}. A fictitious closed boundary is drawn around the scatterer. The fields $\mathbf{E}$ and $\mathbf{H}$ are recorded on this boundary during the simulation. Using the equivalence principle, the scatterer and its surrounding medium are replaced by equivalent electric and magnetic surface currents,
\begin{equation}
    \mathbf{J} = \hat{\mathbf{n}} \times \mathbf{H}, 
    \qquad
    \mathbf{M} = - \hat{\mathbf{n}} \times \mathbf{E},
\end{equation}
where $\hat{\mathbf{n}}$ is the unit outward normal and $\mathbf{E}, \mathbf{H}$ are the electric and magnetic fields on the fictitious boundary.

Since LBM is inherently a time-domain method, we first record the near-field data ($\pmb{\mathcal{E}}, \pmb{\mathcal{H}}$) on the fictitious boundary over one period of the steady-state oscillation. A discrete Fourier transform is then applied to extract the frequency-domain fields at the driving frequency. These frequency-domain fields are used to construct the equivalent surface currents $\mathbf{J}$ and $\mathbf{M}$, which in turn serve as sources for the vector potentials at an observation point in the far-field.

For 2D problems, the vector potentials are expressed as line integrals involving the Hankel function of the second kind,
\begin{subequations}
    \begin{equation}
        \mathbf{A}(\boldsymbol{r}) = -j \frac{3\mu_r}{4} \oint\limits_{L} 
        \mathbf{J}(\boldsymbol{r}') H^{(2)}_{0}(k|\boldsymbol{r} - \boldsymbol{r}'|)\, d\ell',
    \end{equation}
    \begin{equation}
        \mathbf{F}(\boldsymbol{r}) = -j \frac{3\epsilon_r}{4} \oint\limits_{L} 
        \mathbf{M}(\boldsymbol{r}') H^{(2)}_{0}(k|\boldsymbol{r} - \boldsymbol{r}'|)\, d\ell',
    \end{equation}
\end{subequations}
where $\boldsymbol{r}$ is the far-field observation point, $\boldsymbol{r}'$ is the source (surface currents) location, $L$ is the integration contour, $j = \sqrt{-1}$ denotes the imaginary unit, $k = 2 \pi / \lambda$ is the wavenumber and $H_0^{(2)}$ denotes the zeroth-order Hankel function of second kind.

For 3D problems, the corresponding vector potentials are obtained from surface integrals given below,
\begin{subequations}
    \begin{equation}
        \mathbf{A}(\mathbf{r}) = 3\mu_r \oint_{S} \mathbf{J}(\mathbf{r}')
       \frac{e^{-jk|\mathbf{r} - \mathbf{r}'|}}{4\pi |\mathbf{r} - \mathbf{r}'|} ds',
    \end{equation}
    \begin{equation}
        \mathbf{F}(\mathbf{r}) = 3\epsilon_r \oint_{S} \mathbf{M}(\mathbf{r}')
       \frac{e^{-jk|\mathbf{r} - \mathbf{r}'|}}{4\pi |\mathbf{r} - \mathbf{r}'|} ds'.
    \end{equation}
\end{subequations}

Finally, the radiated fields are computed from these vector potentials using
\begin{subequations}
    \begin{equation}
        \mathbf{E}(\boldsymbol{r}) = -j \omega \left[ \mathbf{A} + \frac{1}{k^2} \nabla (\nabla \cdot \mathbf{A}) \right] 
        - \frac{1}{3\epsilon_r} \nabla \times \mathbf{F},
    \end{equation}
    \begin{equation}
        \mathbf{H}(\boldsymbol{r}) = -j \omega \left[ \mathbf{F} + \frac{1}{k^2} \nabla (\nabla \cdot \mathbf{F}) \right] 
        + \frac{1}{3\mu_r} \nabla \times \mathbf{A},
    \end{equation}
\end{subequations}
where $\omega$ denotes the angular frequency of the incident wave, and the factor of 3 arises from the scaling used in the recovered Maxwell’s equations, as discussed earlier.
This NTFF procedure provides a rigorous way to propagate the finite-domain LBM results to the far-field region. By transforming the recorded near-fields into equivalent surface currents and applying integral representations of vector potentials, one obtains consistent definitions of scattering intensities and radiation patterns that can be directly compared with analytical or semi-analytical benchmarks.


\section{Results and Discussion}\label{sec:results}

In this section, we present the LBM results and compare them with available
analytical or semi-analytical solutions in order to validate the proposed
framework. We begin with the fundamental problem of a plane wave normally incident on a planar interface, considering both dielectric and magnetic walls. The LBM predictions for the normalized reflected and transmitted electric fields are compared with analytical expressions over a wide range of dielectric constants and relative permeabilities.

We then consider 2D scatterers, specifically infinitely long circular cylinders of both PEC and dielectric type, for which exact solutions are provided by Lorenz--Mie theory. The angular scattering intensity is computed using both LBM and Lorenz--Mie theory over a broad range of size-to-wavelength ratios. Additional comparisons are performed for dielectric cylinders at fixed size-to-wavelength ratio while varying the dielectric constant to assess accuracy across different material contrasts.

Motivated by the geometry of ice crystals, we further investigate scattering from a regular hexagonal dielectric cylinder of infinite length. For this configuration, LBM results are benchmarked against solutions obtained from the DMF over a wide range of size-to-wavelength ratios.

Finally, we extend the analysis to 3D scatterers by considering a dielectric sphere, where the far-field scattering intensity obtained with LBM is validated against exact Lorenz--Mie solutions.

\subsection{Reflection and refraction at a planar interface}\label{sec:plane_wall}

A fundamental benchmark for electromagnetic wave scattering is the interaction of a plane wave with a flat dielectric or magnetic boundary. We consider a plane wave normally incident from vacuum onto a semi-infinite medium characterized by dielectric constant $\varepsilon_r$ and relative permeability $\mu_r$. Both the vacuum and the medium are treated as semi-infinite to eliminate secondary reflections, and the interface is modeled as a sharp discontinuity. The normalized reflected and transmitted electric fields are computed as functions of $\varepsilon_r$ and $\mu_r$ and compared directly with analytical solutions.

\begin{figure*}[!htb]
\centering
\subfigure[Varying dielectric constant $\varepsilon_r$ ($\mu_r = 1$)]{\includegraphics[width=0.48\textwidth]{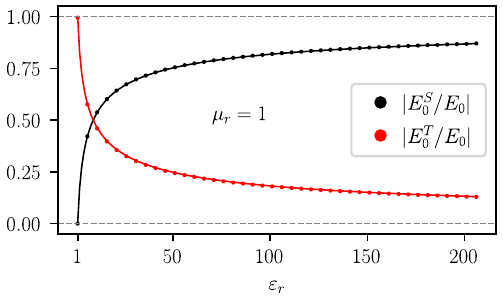}}
\subfigure[Varying relative permeability $\mu_r$ ($\varepsilon_r = 1$)]{\includegraphics[width=0.48\textwidth]{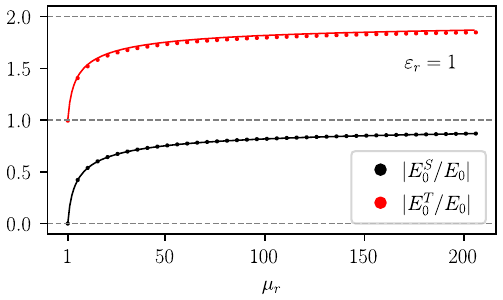}}
\caption{Normalized reflected (black) and transmitted (red) electric field amplitudes at a planar interface (1D) for normal incidence. (a) Variation with dielectric constant $\varepsilon_r$ ($\mu_r = 1$). (b) Variation with relative permeability $\mu_r$ ($\varepsilon_r = 1$). Solid lines denote analytical solutions and markers denote LBM results.}
\label{fig:plane_wall}
\end{figure*}

For normal incidence, the amplitude ratios of the reflected and transmitted electric fields relative to the incident field are given by \cite{griffiths2005introduction}:
\begin{equation} \label{eq:ref_trans}
\frac{E_0^{S}}{E_0} = \frac{\sqrt{\mu_r} - \sqrt{\varepsilon_r}}{\sqrt{\mu_r} + \sqrt{\varepsilon_r}}, \quad
\frac{E_0^{T}}{E_0} = \frac{2 \sqrt{\mu_r}}{\sqrt{\mu_r} + \sqrt{\varepsilon_r}},
\end{equation}
where $E_0$, $E_0^S$, and $E_0^T$ denote the amplitudes of the incident, reflected, and transmitted electric fields, respectively.

The computational grid is resolved with $\lambda_{\text{wall}} / \Delta x = 20$, where $\lambda_{\text{wall}} = \lambda / \sqrt{\mu_r \varepsilon_r}$ is the wavelength inside the medium, $\lambda$ is the free-space wavelength, and $\Delta x$ is the lattice spacing. This resolution ensures accuracy over the full range of material parameters considered ($1 \leq \varepsilon_r \leq 200$ and $1 \leq \mu_r \leq 200$). Two representative cases are examined:  
(1) fixing $\mu_r = 1$ while varying $\varepsilon_r$ from 1 to 200, and  
(2) fixing $\varepsilon_r = 1$ while varying $\mu_r$ from 1 to 200.

For each case, the normalized reflected and transmitted electric fields are computed using the LBM and benchmarked against analytical predictions. Figure~\ref{fig:plane_wall} illustrates the comparisons, which exhibit very close agreement: deviations remain close to $1\%$ even for high material contrasts. This validates the accuracy and robustness of the LBM in modeling wave interactions at planar dielectric and magnetic interfaces.

With this 1D benchmark established, we next extend the analysis to 2D scattering from infinitely long circular and hexagonal cylinders, followed by 3D scattering from a dielectric sphere. In both the 2D and 3D cases, we restrict attention to non-magnetic media by setting $\mu_r = 1$.


\subsection{Scattering from circular cylinders of infinite length}

We consider both PEC and dielectric 2D circular cylinders of radius $a$, illuminated by a normally incident TM plane wave of free-space wavelength $\lambda$. The angular scattering intensity $|S(\theta)|^2$ is computed using the LBM and compared directly with the corresponding analytical Lorenz--Mie solutions. Representative values of the size-to-wavelength ratio $a/\lambda = 0.1, 1$ and $10$ are examined, spanning small, intermediate, and large scattering configurations. For dielectric cylinders, a dielectric constant of $\varepsilon_r = 2$ is considered unless otherwise specified.

\subsubsection{Scattering from 2D PEC circular cylinders}

\begin{figure*}[!htb]
    \centering
    \subfigure[$a / \lambda = 0.1$]{\includegraphics[width=0.32\textwidth]{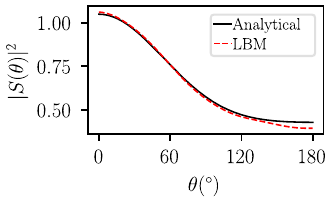}}
    \subfigure[$a / \lambda = 1$]{\includegraphics[width=0.32\textwidth]{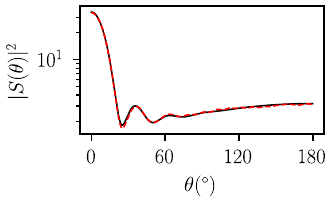}}
    \subfigure[$a / \lambda = 10$]{\includegraphics[width=0.32\textwidth]{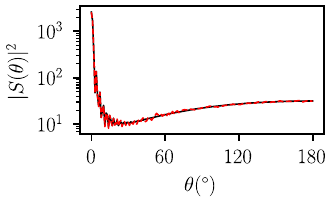}}
    \subfigure[$a / \lambda = 0.1$]{\includegraphics[width=0.32\textwidth]{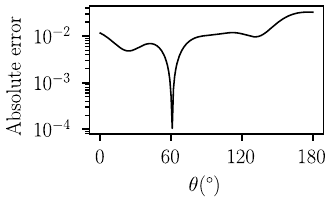}}
    \subfigure[$a / \lambda = 1$]{\includegraphics[width=0.32\textwidth]{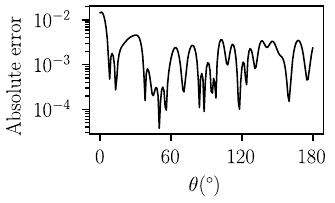}}
    \subfigure[$a / \lambda = 10$]{\includegraphics[width=0.32\textwidth]{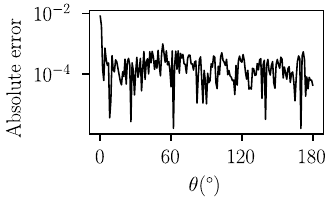}}
    \caption{Angular scattering intensity $|S(\theta)|^2$ for a 2D PEC circular cylinder at different size-to-wavelength ratios. (a--c) Comparison between analytical (solid) and LBM (dashed) results for $a/\lambda = 0.1, 1,$ and $10$. (d--f) Corresponding normalized absolute error.}
    \label{fig:SW_PEC_cylinder}
\end{figure*}

We first consider scattering from an infinitely long PEC circular cylinder under normally incident TM plane wave illumination. The angular scattering intensity $|S(\theta)|^2$ obtained using the LBM is compared with the analytical Lorenz--Mie solution. The top row of Fig.~\ref{fig:SW_PEC_cylinder} presents the comparison. In all three cases, the LBM predictions are in close agreement with the analytical results over the full angular range $0^\circ \le \theta \le 180^\circ$. The method closely captures the forward-scattering peak, side lobes, and the increasingly oscillatory interference structure observed as the size-to-wavelength ratio increases.

To assess the local deviation, the bottom row of Fig.~\ref{fig:SW_PEC_cylinder} shows the absolute error normalized by the maximum value of the analytical reference intensity, defined as

\begin{equation} \label{eq:abs_error}
\text{Normalized absolute error} =
\frac{\left|\, |S(\theta)|^2_{\mathrm{LBM}} - |S(\theta)|^2_{\mathrm{ref}} \,\right|}
{\max_{\theta}\left(|S(\theta)|^2_{\mathrm{ref}}\right)}.
\end{equation}

For $a/\lambda = 0.1$, the angular scattering intensity is of order unity, and the normalized absolute error is of order $10^{-2}$ or lower. For $a/\lambda = 1$, the maximum scattering intensity is of order $10^{1}$ (approximately 30), while the normalized absolute error remains below $10^{-2}$ over most angles, with larger deviations near the forward-scattering region. For $a/\lambda = 10$, the maximum scattering intensity increases to approximately $2.6 \times 10^{3}$, and the normalized absolute error is generally below $10^{-3}$, with larger deviations again confined to the forward-scattering region where the intensity attains its peak value.

To quantify the overall agreement, we compute the normalized root-mean-square (RMS) error over all discrete angular sampling points,

\begin{equation} \label{eq:rms_error}
\text{Normalized RMS} = \frac{1}{{\max_{\theta}\left(|S(\theta)|^2_{\mathrm{ref}}\right)}}
\sqrt{\frac{1}{N} \sum_{i=1}^{N}
\left(
|S(\theta_i)|^2_{\mathrm{LBM}} -
|S(\theta_i)|^2_{\mathrm{ref}}
\right)^2 },
\end{equation}
where $N$ denotes the total number of angular samples. The corresponding RMS values, together with the minimum and maximum values of the angular scattering intensity, are summarized in Table~\ref{table:cylinder_RMS_combined}. The normalized RMS error decreases with increasing size-to-wavelength ratio and remains below $1.5\times10^{-2}$ in all cases, indicating good agreement between the LBM predictions and the analytical solutions.

For the simulations of 2D PEC circular cylinders, the spatial resolution is chosen such that
\[
\min\{a/\Delta x, \lambda/\Delta x\} = 50.
\]
Accordingly, for $a/\lambda \le 1$, the cylinder radius is resolved with 50 grid points, whereas for $a/\lambda > 1$, the incident wavelength is resolved with 50 grid points. The computational domain is taken as a square of side length $L$, with $L/a = 20$ for $a/\lambda < 1$, $L/a = 10$ for $a/\lambda = 1$, and $L/a = 4$ for $a/\lambda > 1$. The corresponding grid parameters and computation times are summarized in Table~\ref{table:cylinder_simulation_parameters}. It is worth noting that the relatively larger computation times observed for smaller values of $a/\lambda$ are primarily due to the larger computational domain sizes employed in these cases, which increase both the total number of lattice points and the time required to reach steady state.


\subsubsection{Scattering from 2D dielectric circular cylinders}

\begin{figure*}[!htb]
    \centering
    \subfigure[$a / \lambda = 0.1$]{\includegraphics[width=0.32\textwidth]{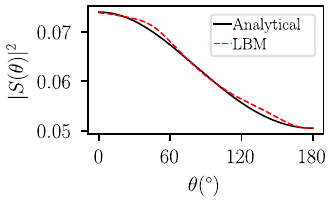}}
    \subfigure[$a / \lambda = 1$]{\includegraphics[width=0.32\textwidth]{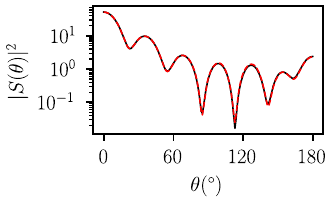}}
    \subfigure[$a / \lambda = 10$]{\includegraphics[width=0.32\textwidth]{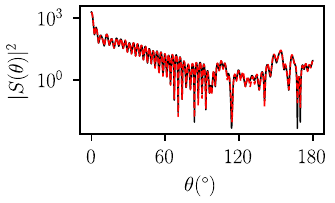}}
    \subfigure[$a / \lambda = 0.1$]{\includegraphics[width=0.32\textwidth]{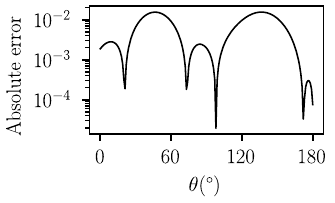}}
    \subfigure[$a / \lambda = 1$]{\includegraphics[width=0.32\textwidth]{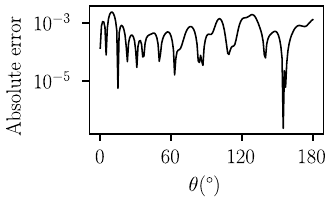}}
    \subfigure[$a / \lambda = 10$]{\includegraphics[width=0.32\textwidth]{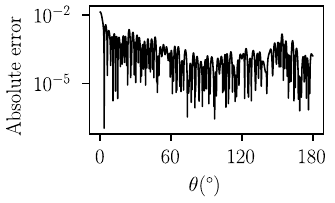}}
    \caption{Angular scattering intensity $|S(\theta)|^2$ for a 2D dielectric circular cylinder with $\varepsilon_r = 2$. (a--c) Comparison between analytical (solid) and LBM (dashed) results for $a/\lambda = 0.1, 1,$ and $10$. (d--f) Corresponding normalized absolute error.}
    \label{fig:SW_er2_cylinder}
\end{figure*}

We next consider a 2D dielectric circular cylinder with dielectric constant $\varepsilon_r = 2$ under normally incident TM plane wave illumination. The angular scattering intensity $|S(\theta)|^2$ computed using the LBM is compared with the analytical Lorenz–Mie solution. As shown in the top row of Fig.~\ref{fig:SW_er2_cylinder}, the LBM predictions closely follow the analytical results over the full angular range. The main scattering features—including the forward-scattering peak, side lobes, and interference oscillations—are accurately captured. Minor discrepancies appear only near a few sharp minima for $a/\lambda = 10$, where the scattering intensity varies rapidly with angle.

The bottom row of Fig.~\ref{fig:SW_er2_cylinder} shows the normalized absolute error, defined as in the PEC case. For all values of $a/\lambda$, the error remains below $10^{-2}$, with most values below $10^{-3}$. Slightly larger errors are observed for $a/\lambda = 0.1$, where the scattering intensity itself is of order $10^{-2}$, as well as near sharp minima for $a/\lambda = 1$ and in the forward-scattering region for $a/\lambda = 10$. The corresponding normalized RMS errors, summarized in Table~\ref{table:cylinder_RMS_combined}, remain below $10^{-2}$ for all cases.

A key numerical consideration for dielectric cylinders is the reduction of the internal wavelength,
\[
\lambda_{\varepsilon_r} = \frac{\lambda}{\sqrt{\varepsilon_r}},
\]
which necessitates finer spatial resolution to accurately resolve wave propagation inside the cylinder. Accordingly, for 2D dielectric circular cylinders with $\varepsilon_r = 2$, the grid resolution is chosen such that
\[
\min\{a/\Delta x, \lambda/\Delta x\} = 100,
\]
i.e., twice the resolution used for PEC simulations. For $a/\lambda \le 1$, the cylinder radius is resolved with 100 grid points, whereas for $a/\lambda > 1$, the incident wavelength is resolved with 100 grid points.

The computational domain is taken as a square of side length $L$, with $L/a = 20$ for $a/\lambda < 1$, $L/a = 10$ for $a/\lambda = 1$, and $L/a = 4$ for $a/\lambda > 1$. The corresponding domain sizes, grid resolutions, and computation times are summarized in Table~\ref{table:cylinder_simulation_parameters}.

\begin{figure*}[!t]
    \centering
    \subfigure[$\varepsilon_r = 5$]{\includegraphics[width=0.32\textwidth]{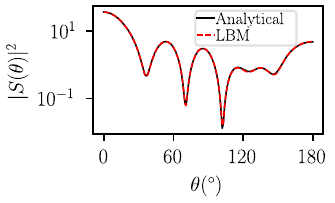}}
    \subfigure[$\varepsilon_r = 10$]{\includegraphics[width=0.32\textwidth]{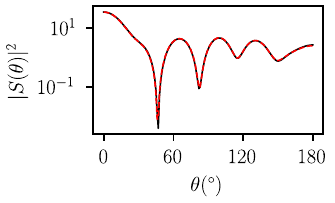}}
    \subfigure[$\varepsilon_r = 20$]{\includegraphics[width=0.32\textwidth]{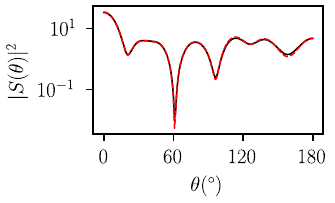}}
    \subfigure[$\varepsilon_r = 5$]{\includegraphics[width=0.32\textwidth]{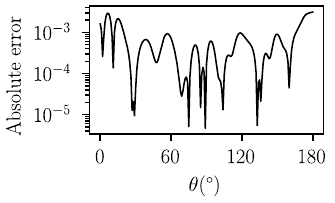}}
    \subfigure[$\varepsilon_r = 10$]{\includegraphics[width=0.32\textwidth]{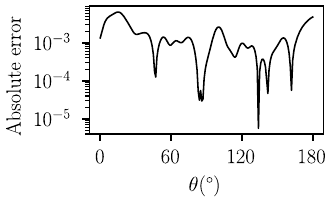}}
    \subfigure[$\varepsilon_r = 20$]{\includegraphics[width=0.32\textwidth]{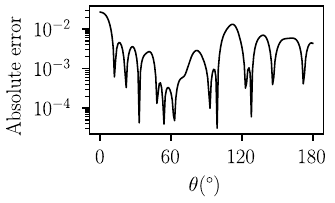}}
    \caption{Angular scattering intensity $|S(\theta)|^2$ for a 2D dielectric circular cylinder at $a/\lambda = 1$. (a--c) Comparison between analytical (solid) and LBM (dashed) results for $\varepsilon_r = 5, 10,$ and $20$. (d--f) Corresponding normalized absolute error.}
    \label{fig:SW_ratio1_cylinder}
\end{figure*}

\begin{table}[!b]
\centering
\caption{Minimum and maximum values of the angular scattering intensity $|S(\theta)|^2$ and the corresponding normalized RMS error between LBM and analytical Lorenz--Mie solutions for 2D circular cylinders. The RMS error is computed over the full angular range.}
\begin{tabular}{ccccc}
\hline
\textbf{Case} & \textbf{Parameter} & \textbf{$\min_{\theta}(|S(\theta)|^2)$} & \textbf{$\max_{\theta}(|S(\theta)|^2)$} & \textbf{Normalized RMS error} \\
\hline
\multirow{3}{*}{PEC}
 & $a/\lambda = 0.1$ & 0.428 & 1.05 & $1.5 \times 10^{-2}$ \\
 & $a/\lambda = 1$    & 1.82 & 34.2 & $3.0 \times 10^{-3}$ \\
 & $a/\lambda = 10$   & 8.81 & $2.6 \times 10^{3}$ & $7.2 \times 10^{-4}$ \\
\hline
\multirow{3}{*}{Dielectric ($\varepsilon_r=2$)}
 & $a/\lambda = 0.1$ & 0.050 & 0.074 & $8.4 \times 10^{-3}$ \\
 & $a/\lambda = 1$    & 0.016 & 51.7 & $7.8 \times 10^{-4}$ \\
 & $a/\lambda = 10$   & 0.004 & $2.0 \times 10^3$ & $1.3 \times 10^{-3}$ \\
\hline
\multirow{3}{*}{$a/\lambda=1$}
 & $\varepsilon_r = 5$  & 0.012 & 37.0 & $1.0 \times 10^{-3}$ \\
 & $\varepsilon_r = 10$ & 0.004 & 34.9 & $2.3 \times 10^{-3}$ \\
 & $\varepsilon_r = 20$ & 0.011 & 33.4 & $6.6 \times 10^{-3}$ \\
\hline
\end{tabular}
\label{table:cylinder_RMS_combined}
\end{table}

Having validated the method across a wide range of size-to-wavelength ratios for both PEC and dielectric 2D circular cylinders, we now assess the performance of the LBM for high dielectric contrast. To this end, we fix the size-to-wavelength ratio at $a/\lambda = 1$ and vary the dielectric constant of the cylinder. Figure~\ref{fig:SW_ratio1_cylinder} shows the results for $\varepsilon_r = 5, 10,$ and $20$. The LBM predictions closely follow the analytical solutions over the full angular range. The corresponding normalized absolute error remains below $10^{-2}$, with most values below $10^{-3}$, and the normalized RMS error (Table~\ref{table:cylinder_RMS_combined}) remains below $7 \times 10^{-3}$ for all cases.

For these simulations, the grid resolution was chosen such that $\lambda_{\varepsilon_r}/\Delta x \approx 50$, with the domain size fixed at $L/a = 10$. The corresponding grid resolutions and computation times are summarized in Table~\ref{table:cylinder_simulation_parameters}. As expected, increasing $\varepsilon_r$ enhances internal reflections within the cylinder \cite{hulst1981light}, thereby extending the simulation time required to reach steady state. Convergence was verified by monitoring the temporal evolution of the total field energy inside the computational domain and ensuring its stabilization over time.

\begin{table}[!htb]
\centering
\caption{Grid resolution, domain size, and computation time for LBM simulations of 2D circular cylinders under different parameters. The table lists $L/a$, $a/\Delta x$, $\lambda/\Delta x$ or $\lambda_{\varepsilon_r}/\Delta x$, the number of threads, and the corresponding simulation time.}
\begin{tabular}{ccccccc}
\hline
\textbf{Case} & \textbf{Parameter} & \textbf{$L/a$} & \textbf{$a/\Delta x$} & \textbf{$\lambda/\Delta x$ or $\lambda_{\varepsilon_r}/\Delta x$} & \textbf{Thread} & \textbf{Time (hrs)} \\
\hline
\multirow{3}{*}{PEC}
 & $a/\lambda = 0.1$ & 20 & 50   & 500  & 20 & 0.51 \\
 & $a/\lambda = 1$    & 10 & 50   & 50   & 20 & 0.07 \\
 & $a/\lambda = 10$   & 4  & 500  & 50   & 20 & 3.8 \\
\hline
\multirow{3}{*}{Dielectric ($\varepsilon_r=2$)}
 & $a/\lambda = 0.1$ & 20 & 100  & 1000 & 14 & 3.6 \\
 & $a/\lambda = 1$    & 10 & 100  & 100  & 14 & 0.33 \\
 & $a/\lambda = 10$   & 4  & 1000 & 100  & 14 & 30.4 \\
\hline
\multirow{3}{*}{$a/\lambda=1$}
 & $\varepsilon_r = 5$  & 10 & 112 & 50 & 20 & 1.2 \\
 & $\varepsilon_r = 10$ & 10 & 158 & 50 & 20 & 3.1 \\
 & $\varepsilon_r = 20$ & 10 & 224 & 50 & 20 & 32.7 \\
\hline
\end{tabular}
\label{table:cylinder_simulation_parameters}
\end{table}

Having established the accuracy of the LBM for canonical 2D circular cylinders, we next assess its performance for non-circular scatterers — specifically, a 2D regular hexagonal dielectric cylinder — where analytical solutions are not available and the DMF serves as the benchmark.


\subsection{Scattering from a hexagonal dielectric cylinder}
\label{sec:hex}

Canonical geometries such as spheres and 2D circular cylinders admit exact analytical solutions, but many physically relevant scatterers exhibit sharp-edged features and faceted surfaces. A particularly important example is the regular hexagonal cylinder, long recognized as a simplified model for atmospheric ice crystals. The presence of flat facets and corners gives rise to strong diffraction and localized field enhancements, making the hexagon a demanding test case for numerical solvers.

\begin{figure*}[!htb]
    \centering
    \subfigure[$a/\lambda = 0.1$.]{\includegraphics[width=0.33\textwidth]{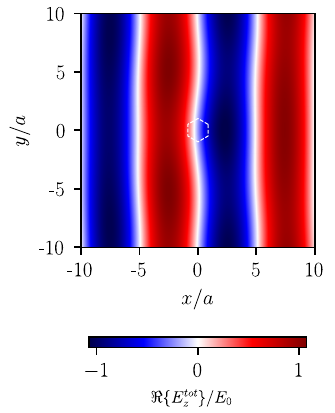}}
    \subfigure[$a/\lambda = 1$.]{\includegraphics[width=0.33\textwidth]{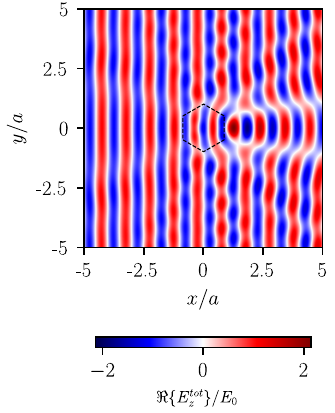}}
    \subfigure[$a/\lambda = 10$.]{\includegraphics[width=0.33\textwidth]{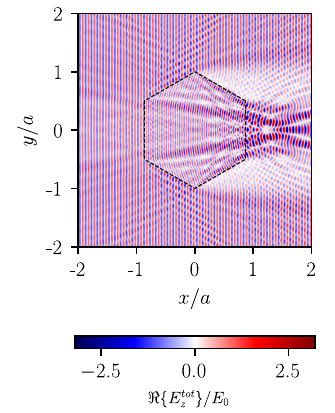}}
    \caption{Real part of the total electric field for a 2D hexagonal dielectric cylinder with $\varepsilon_r = 1.721$. (a--c) Results for $a/\lambda = 0.1, 1,$ and $10$. The dashed line indicates the cylinder boundary.}
    \label{fig:hexagon}
\end{figure*}

We consider a regular hexagonal dielectric cylinder illuminated by a plane wave incident normal to its axis. Both transverse electric (TE) and transverse magnetic (TM) polarizations are analyzed. The dielectric constant of the cylinder is taken as $\varepsilon_r = 1.721$, corresponding to that of ice at visible wavelengths~\cite{liou2016light}.

Figure~\ref{fig:hexagon} shows representative snapshots of the real part of the total electric field obtained from LBM simulations for TM polarization at $a/\lambda = 0.1$, $1$, and $10$. The field distributions illustrate the complex scattering patterns generated by the faceted geometry of the hexagonal cylinder.

\begin{figure*}[!htb]
    \centering
    \subfigure[$a / \lambda = 1 / 10$]{\includegraphics[width=0.32\textwidth]{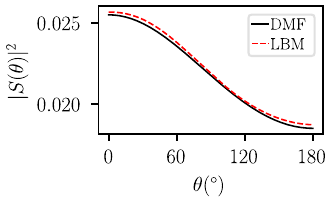}}
    \subfigure[$a / \lambda = 1$]{\includegraphics[width=0.32\textwidth]{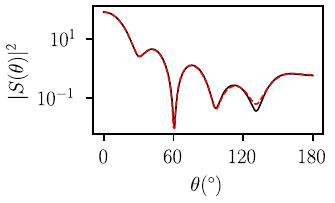}}
    \subfigure[$a / \lambda = 10$]{\includegraphics[width=0.32\textwidth]{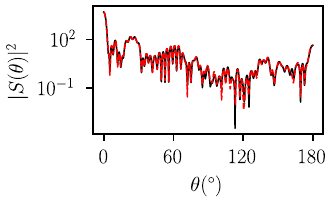}}
    \subfigure[$a / \lambda = 1 / 10$]{\includegraphics[width=0.32\textwidth]{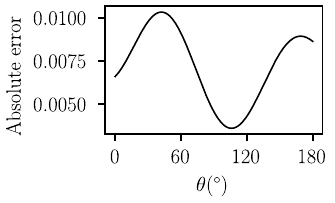}}
    \subfigure[$a / \lambda = 1$]{\includegraphics[width=0.32\textwidth]{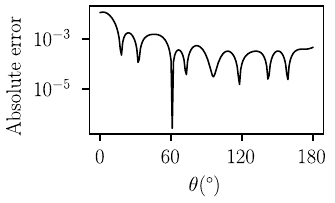}}
    \subfigure[$a / \lambda = 10$]{\includegraphics[width=0.32\textwidth]{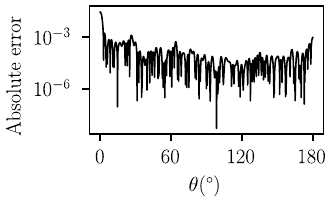}}
    \caption{Angular scattering intensity $|S(\theta)|^2$ for a 2D hexagonal dielectric cylinder with $\varepsilon_r = 1.721$ (TM polarization). (a--c) Comparison between DMF (solid) and LBM (dashed) results for $a/\lambda = 0.1, 1,$ and $10$. (d--f) Corresponding normalized absolute error.}
    \label{fig:hexagon_TM}
\end{figure*}

\begin{figure*}[hbt]
    \centering
    \subfigure[$a / \lambda = 1 / 10$]{\includegraphics[width=0.32\textwidth]{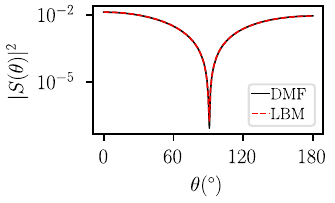}}
    \subfigure[$a / \lambda = 1$]{\includegraphics[width=0.32\textwidth]{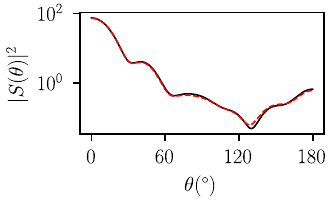}}
    \subfigure[$a / \lambda = 10$]{\includegraphics[width=0.32\textwidth]{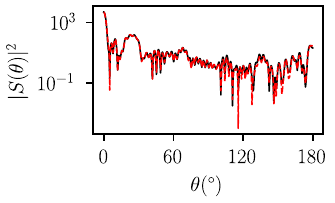}}
    \subfigure[$a / \lambda = 1 / 10$]{\includegraphics[width=0.32\textwidth]{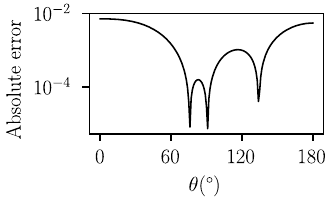}}
    \subfigure[$a / \lambda = 1$]{\includegraphics[width=0.32\textwidth]{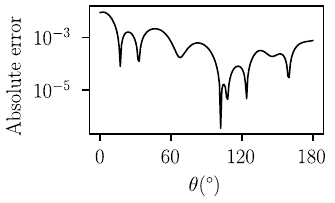}}
    \subfigure[$a / \lambda = 10$]{\includegraphics[width=0.32\textwidth]{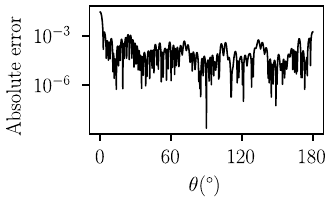}}
    \caption{Angular scattering intensity $|S(\theta)|^2$ for a 2D hexagonal dielectric cylinder with $\varepsilon_r = 1.721$ (TE polarization). (a--c) Comparison between DMF (solid) and LBM (dashed) results for $a/\lambda = 0.1, 1,$ and $10$. (d--f) Corresponding normalized absolute error.}
    \label{fig:hexagon_TE}
\end{figure*}

The angular scattering intensity $|S(\theta)|^2$ computed using the LBM is compared with results obtained using the DMF developed by Rother and Schmidt~\cite{rother1997discretized}, implemented in-house based on their formulation. Figures~\ref{fig:hexagon_TM} and~\ref{fig:hexagon_TE} present the comparisons for TM and TE polarizations, respectively, together with the corresponding normalized absolute errors.

For $a/\lambda = 0.1$, the normalized absolute error remains below $10^{-2}$ for both TM and TE polarizations. For $a/\lambda = 1$ and $10$, the error is mostly below $10^{-3}$, with larger deviations confined to the forward-scattering region where the angular scattering intensity attains its highest values. A quantitative comparison is provided in Table~\ref{table:dielectric_cylinder_hex_RMS}, where the normalized RMS error remains below $8 \times 10^{-3}$ for all cases.

For the LBM simulations, the grid resolution was chosen such that 
$\min\{a/\Delta x,\, \lambda_{\varepsilon_r}/\Delta x\} \approx 40$ for both TM and TE polarization cases, ensuring accurate resolution of both the geometry and the internal field. Here $a$ denotes the radius of the circumscribed circle of the hexagon and $\lambda_{\varepsilon_r}$ is the wavelength inside the dielectric. The computational domain sizes, grid resolutions, and execution times are listed in Table~\ref{table:dielectric_cylinder_hex}.  

For the DMF calculations, the angular coordinate was discretized uniformly with spacing $h_{\theta}=2\pi/N_d$, where $N_d$ is the number of discrete angular points chosen such that none coincides with a vertex. Converged results were obtained using $N_d = 361, 361,$ and $30551$ for $a/\lambda = 0.1, 1,$ and $10$, respectively, and truncation parameters $N_{\text{cut}} = 100, 100,$ and $500$, where $N_{\text{cut}}$ denotes the number of retained multipole terms in the DMF expansion. These parameters were used for both TM and TE polarization cases.

\begin{table}[!htb]
\centering
\caption{Minimum and maximum values of the angular scattering intensity $|S(\theta)|^2$ and the corresponding normalized RMS error between LBM and DMF solutions for hexagonal dielectric cylinders at different size-to-wavelength ratios. The RMS error is computed over the full angular range.}
\begin{tabular}{ccccc}
\hline
\textbf{Polarization} & \textbf{$a/\lambda$} & \textbf{$\min_{\theta}(|S(\theta)|^2)$} & \textbf{$\max_{\theta}(|S(\theta)|^2)$} & \textbf{Normalized RMS error} \\
\hline
\multirow{3}{*}{TM}
 & 0.1 & $1.8 \times 10^{-2}$ & 0.025 & $7.6 \times 10^{-3}$ \\
 & 1    & $9.0 \times 10^{-3}$ & 81.9 & $2.7 \times 10^{-3}$ \\
 & 10   & $3.2 \times 10^{-4}$ & $4.9 \times 10^3$ & $2.3 \times 10^{-3}$ \\
\hline
\multirow{3}{*}{TE}
 & 0.1 & $8.4 \times 10^{-8}$ & 0.013 & $3.8 \times 10^{-3}$ \\
 & 1    & $4.8 \times 10^{-2}$ & 74.3 & $2.1 \times 10^{-3}$ \\
 & 10   & $3.0 \times 10^{-3}$ & $4.9 \times 10^3$ & $2.1 \times 10^{-3}$ \\
\hline
\end{tabular}
\label{table:dielectric_cylinder_hex_RMS}
\end{table}

\begin{table}[!htb]
\centering
\caption{Grid resolution, domain size, and computation time for LBM simulations of hexagonal dielectric cylinders ($\varepsilon_r = 1.721$) at different size-to-wavelength ratios. Columns list $L/a$, $a/\Delta x$, $\lambda_{\varepsilon_r}/\Delta x$, number of threads, and simulation time.}
\begin{tabular}{cccccc}
\hline
\textbf{$a/\lambda$} & \textbf{$L/a$} & \textbf{$a/\Delta x$} & \textbf{$\lambda_{\varepsilon_r}/\Delta x$} & \textbf{Thread} & \textbf{Time (hrs)} \\
\hline
0.1 & 20    & 40   & 305  & 10 & 0.05   \\
1    & 10    & 52   & 40   & 10 & 0.04   \\
10   & 4     & 525  & 40   & 10 & 2.6  \\
\hline
\label{table:dielectric_cylinder_hex}
\end{tabular}
\end{table}

The complementary nature of the two methods is noteworthy: DMF is semi-analytical, discretizing only the angular dependence while solving the radial part exactly, whereas LBM is a fully time-domain, grid-based solver. Their close agreement demonstrates not only the accuracy of LBM for sharp-edged scatterers but also its flexibility for complex geometries where semi-analytical approaches become intractable.

\subsection{Scattering from a Dielectric Sphere}\label{sec:sphere}

We now extend the analysis to the three-dimensional case of electromagnetic scattering by a dielectric sphere with dielectric constant $\varepsilon_r = 2$. A plane electromagnetic wave is incident along the $z$-axis, with the electric field polarized along the $x$-axis and the magnetic field along the $y$-axis, as illustrated schematically in Fig.~\ref{fig:schemaic_3D}. The angular scattering intensity $|S(\theta,\phi)|^2$ obtained from the LBM simulations is compared with the corresponding analytical Lorenz--Mie solution.

\begin{figure}[!htb]
\centering
\begin{tikzpicture}[scale=1.5,
    axis/.style={very thick,->,>=Stealth},
    vector/.style={-Stealth,thick},
    dashedline/.style={dashed,thick}
]

\draw[axis] (0,0,0) -- (1.4,0,0) node[right] {$y$};
\draw[axis] (0,0,0) -- (0,1.4,0) node[above] {$z$};
\draw[axis] (0,0,0) -- (0,0,2.5) node[left] {$x$};

\shade[ball color=gray!30, opacity=0.9] (0,0,0) circle (1);

\pgfmathsetmacro{\Theta}{50}  
\pgfmathsetmacro{\Phi}{50}    
\pgfmathsetmacro{\r}{3}
\pgfmathsetmacro{\x}{\r*sin(\Theta)*cos(\Phi)}
\pgfmathsetmacro{\y}{\r*sin(\Theta)*sin(\Phi)}
\pgfmathsetmacro{\z}{\r*cos(\Theta)}

\coordinate (P) at (\y,\z,\x);

\draw[vector] (0,0,0) -- (-0.4,1,1) node[midway,above] {$a$};

\fill (P) circle(0.6pt);
\draw[vector] (0,0,0) -- (P) node[pos=0.6, yshift=6pt] {$r$};
\draw[dashedline] (0,0,0) -- (\y,0,\x);
\draw[dashedline] (\y,0,\x) -- (P);

\draw[->] (0,0.5,0) arc[start angle=90,end angle=\Theta,radius=0.5] node[midway, above] {$\theta$};

\draw[->] (0,0,0.5) arc[start angle=-2*\Phi,end angle=-1.09*\Phi,radius=0.5] node[midway, below] {$\phi$};

\draw[vector] (0,-1.3,0) -- (0,-1.3,0.5) node[left] {${\bf E}$};
\draw[vector] (0,-1.3,0) -- (0.3,-1.3,0) node[right] {${\bf H}$};

\draw[-Stealth,thin] (0,-1.5,0) -- (0,-1.1,0);

\end{tikzpicture}
\caption{Spherical coordinate system for scattering by a dielectric sphere. The incident wave propagates along the $z$-axis with the electric and magnetic fields polarized along the $x$- and $y$-axes, respectively.}
\label{fig:schemaic_3D}
\end{figure}

\begin{figure*}[!htb]
    \centering

    \subfigure[$a/ \lambda = 0.1$]{
    \begin{tikzpicture}
        \node[anchor=south west,inner sep=0] (img) {\includegraphics[width=0.48\textwidth]{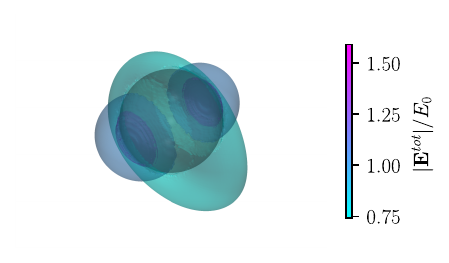}};
        \begin{scope}[x={(img.south east)},y={(img.north west)}]
            \begin{scope}[shift={(0.12,0.18)},scale=0.12,thick,->,>=stealth]
                \draw (0,0) -- (0,1.0) node[above] {$z$};
                \draw (0,0) -- ({-0.7*cos(45)},{-0.7*sin(45)}) node[below left] {$x$};
                \draw (0,0) -- ({0.7*cos(45)},{-0.7*sin(45)}) node[below right] {$y$};
            \end{scope}
        \end{scope}
    \end{tikzpicture}
    }
    \subfigure[$a/ \lambda = 1$]{
    \begin{tikzpicture}
        \node[anchor=south west,inner sep=0] (img) {\includegraphics[width=0.48\textwidth]{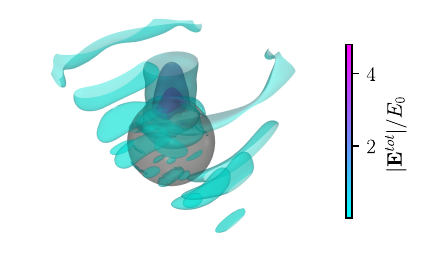}};
        \begin{scope}[x={(img.south east)},y={(img.north west)}]
            \begin{scope}[shift={(0.12,0.18)},scale=0.12,thick,->,>=stealth]
                \draw (0,0) -- (0,1.0) node[above] {$z$};
                \draw (0,0) -- ({-0.7*cos(45)},{-0.7*sin(45)}) node[below left] {$x$};
                \draw (0,0) -- ({0.7*cos(45)},{-0.7*sin(45)}) node[below right] {$y$};
            \end{scope}
        \end{scope}
    \end{tikzpicture}
    }

    \caption{Magnitude of the total electric field around a dielectric sphere with $\varepsilon_r = 2$. (a,b) Results for $a/\lambda = 0.1$ and $1$.}
    \label{fig:snap_sphere}
\end{figure*}

Figure~\ref{fig:snap_sphere} presents representative snapshots of the normalized magnitude of the total electric field (incident plus scattered) around the dielectric sphere for two size-to-wavelength ratios, $a/\lambda = 0.1$ and $a/\lambda = 1$. The field magnitude is normalized by the incident electric field amplitude $E_0$. These snapshots illustrate the spatial distribution of the electric field in the vicinity of the sphere for the two representative size-to-wavelength ratios.

\begin{figure*}[!htb]
    \centering
    \subfigure[$\phi = 0^{\circ}$]{\includegraphics[width=0.33\textwidth]{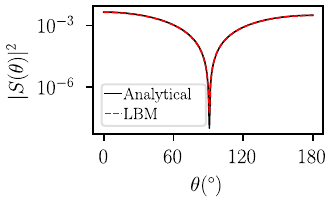}}
    \subfigure[$\phi = 45^{\circ}$]{\includegraphics[width=0.33\textwidth]{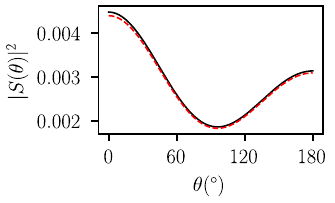}}
    \subfigure[$\phi = 90^{\circ}$]{\includegraphics[width=0.33\textwidth]{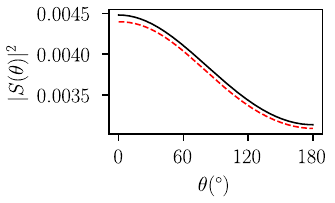}}
    \subfigure[$\phi = 0^{\circ}$]{\includegraphics[width=0.33\textwidth]{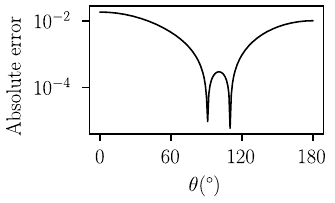}}
    \subfigure[$\phi = 45^{\circ}$]{\includegraphics[width=0.33\textwidth]{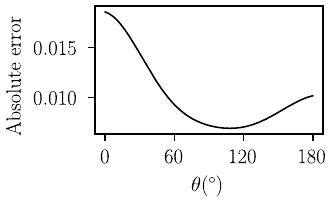}}
    \subfigure[$\phi = 90^{\circ}$]{\includegraphics[width=0.33\textwidth]{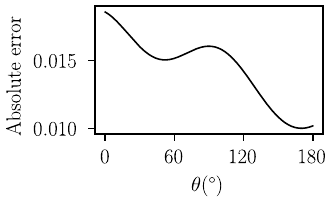}}
    \caption{Angular scattering intensity $|S(\theta,\phi)|^2$ for a dielectric sphere with $\varepsilon_r = 2$ at $a/\lambda = 0.1$. (a--c) Comparison between analytical (solid) and LBM (dashed) results for $\phi = 0^\circ, 45^\circ,$ and $90^\circ$. (d--f) Corresponding normalized absolute error.}
    \label{fig:sphere_01}
\end{figure*}

\begin{figure*}[!htb]
    \centering
    \subfigure[$\phi = 0^{\circ}$]{\includegraphics[width=0.33\textwidth]{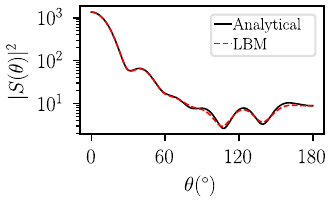}}
    \subfigure[$\phi = 45^{\circ}$]{\includegraphics[width=0.33\textwidth]{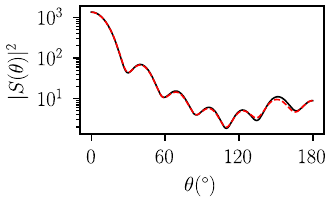}}
    \subfigure[$\phi = 90^{\circ}$]{\includegraphics[width=0.33\textwidth]{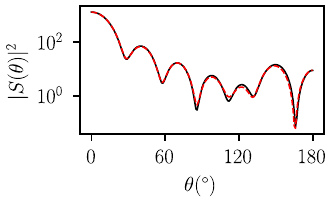}}
    \subfigure[$\phi = 0^{\circ}$]{\includegraphics[width=0.33\textwidth]{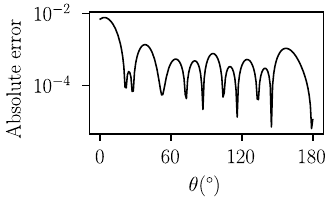}}
    \subfigure[$\phi = 45^{\circ}$]{\includegraphics[width=0.33\textwidth]{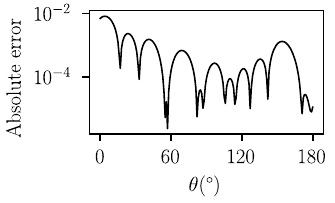}}
    \subfigure[$\phi = 90^{\circ}$]{\includegraphics[width=0.33\textwidth]{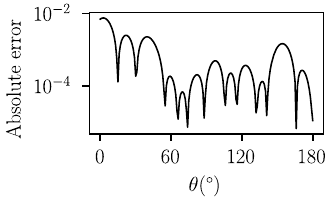}}
    \caption{Angular scattering intensity $|S(\theta,\phi)|^2$ for a dielectric sphere with $\varepsilon_r = 2$ at $a/\lambda = 1$. (a--c) Comparison between analytical (solid) and LBM (dashed) results for $\phi = 0^\circ, 45^\circ,$ and $90^\circ$. (d--f) Corresponding normalized absolute error.}
    \label{fig:sphere_1}
\end{figure*}

\begin{figure*}[!htb]
    \centering
    \subfigure[$\phi = 0^{\circ}$]{\includegraphics[width=0.33\textwidth]{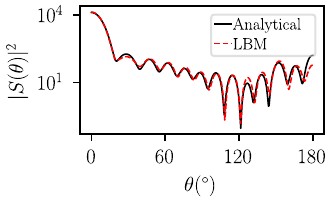}}
    \subfigure[$\phi = 45^{\circ}$]{\includegraphics[width=0.33\textwidth]{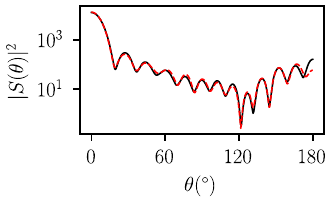}}
    \subfigure[$\phi = 90^{\circ}$]{\includegraphics[width=0.33\textwidth]{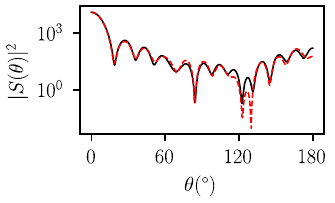}}
    \subfigure[$\phi = 0^{\circ}$]{\includegraphics[width=0.33\textwidth]{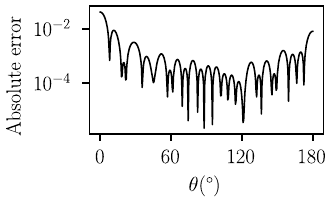}}
    \subfigure[$\phi = 45^{\circ}$]{\includegraphics[width=0.33\textwidth]{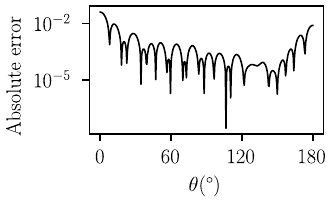}}
    \subfigure[$\phi = 90^{\circ}$]{\includegraphics[width=0.33\textwidth]{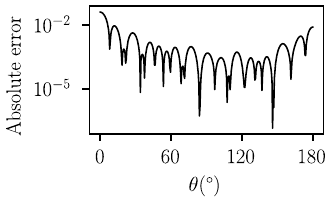}}
    \caption{Angular scattering intensity $|S(\theta,\phi)|^2$ for a dielectric sphere with $\varepsilon_r = 2$ at $a/\lambda = 2$. (a--c) Comparison between analytical (solid) and LBM (dashed) results for $\phi = 0^\circ, 45^\circ,$ and $90^\circ$. (d--f) Corresponding normalized absolute error.}
    \label{fig:sphere_2}
\end{figure*}

The results for dielectric spheres at size-to-wavelength ratios $a/\lambda = 0.1, 1,$ and $2$ are shown in Figs.~\ref{fig:sphere_01}, \ref{fig:sphere_1}, and \ref{fig:sphere_2}, where the corresponding normalized absolute errors are also presented. The results are evaluated for three representative azimuthal angles, $\phi = 0^\circ$, $45^\circ$, and $90^\circ$, corresponding to field orientations aligned with the incident electric field, an intermediate configuration, and orthogonal to it, respectively.

For the smaller size-to-wavelength ratios ($a/\lambda = 0.1$ and $1$), the LBM predictions closely follow the analytical results over the full angular range, with only minor deviations visible in the top rows of Figs.~\ref{fig:sphere_01} and~\ref{fig:sphere_1}. The normalized absolute error (bottom rows) remains below $10^{-1}$ for $a/\lambda = 0.1$, while the corresponding scattering intensity is of order $10^{-3}$. For $a/\lambda = 1$, the error remains below $10^{-2}$ and below $10^{-3}$ over most angles, with slightly larger values near the forward-scattering region. The normalized RMS error for both cases remains below $10^{-2}$ (Table~\ref{table:sphere_RMS}).

For the larger size-to-wavelength ratio $a/\lambda = 2$, the scattering pattern exhibits rapid angular oscillations and increased sensitivity to spatial resolution. While the LBM captures the overall structure of the scattering pattern, noticeable discrepancies appear, particularly in the backscattering direction ($\theta = 180^\circ$) and in localized angular regions (e.g., near $\theta = 120^\circ$ for $\phi = 90^\circ$, see Fig.~\ref{fig:sphere_2}(c)).

The relative error in the backscattering direction is approximately $50\%$, with localized peaks that reach higher values in regions where the scattered intensity is small. This behavior is also reflected in the increased normalized RMS error (Table~\ref{table:sphere_RMS}), which is significantly larger than for smaller size-to-wavelength ratios. These discrepancies are attributed to insufficient spatial resolution to fully resolve the rapidly varying angular features at this size-to-wavelength ratio. In particular, deviations are most pronounced in the backscattering direction, where accurate resolution requires finer discretization. Increasing the resolution (e.g., $\lambda_{\varepsilon_r}/\Delta x \approx 100$) would reduce these errors, but at a significantly higher computational cost in terms of memory and runtime. Improving accuracy in such cases is therefore identified as an important direction for future work.

\begin{table}[!htb]
\centering
\caption{Minimum and maximum values of the angular scattering intensity $|S(\theta,\phi)|^2$ and the corresponding normalized root-mean-square (RMS) error between LBM and analytical Lorenz--Mie solutions for a dielectric sphere at different size-to-wavelength ratios $a/\lambda$ and azimuthal angles $\phi$. The RMS error is computed over the full angular range.}
\begin{tabular}{ccccc}
\hline
\textbf{$a/\lambda$} & \textbf{$\phi^{\circ}$} & \textbf{$\min_{\theta}(|S|^2)$} & \textbf{$\max_{\theta}(|S|^2)$} & \textbf{Normalized RMS error} \\
\hline
\multirow{3}{*}{$0.1$} 
 & 0   & $1.0 \times10^{-8}$ & \multirow{3}{*}{$4.0 \times 10^{-3}$}     & $8.8 \times 10^{-3}$ \\
 & 45  & $1.0 \times 10^{-3}$            &                            & $1.1 \times 10^{-2}$ \\
 & 90  & $3.0 \times 10^{-3}$            &                            & $1.5 \times 10^{-2}$ \\
\hline
\multirow{3}{*}{1} 
 & 0   & 2.7 & \multirow{3}{*}{$1.4 \times 10^3$}  & $1.9 \times 10^{-3}$ \\
 & 45  & 1.9 &                              & $2.0 \times 10^{-3}$ \\
 & 90  & 0.07 &                              & $1.8 \times 10^{-3}$ \\
\hline
\multirow{3}{*}{2} 
 & 0   & 0.09 & \multirow{3}{*}{$1.3 \times 10^4$} & 0.18 \\
 & 45  & 0.34 &                              & 0.18 \\
 & 90  & 0.18 &                              & 0.18 \\
\hline
\end{tabular}
\label{table:sphere_RMS}
\end{table}

For dielectric spheres with $\varepsilon_r = 2$, the grid resolution was selected based on both the particle size and the internal wavelength. For the smallest size-to-wavelength ratio ($a/\lambda = 0.1$), the sphere radius was resolved with $a/\Delta x = 30$. For the larger size-to-wavelength ratios ($a/\lambda = 1$ and $2$), the resolution was chosen such that the internal wavelength was adequately resolved, with $\lambda_{\varepsilon_r}/\Delta x \approx 50$. This ensures sufficient resolution of both the particle geometry and the internal electromagnetic fields.

\begin{table}[!htb]
\centering
\caption{Grid resolution, domain size, and computation time for LBM simulations of dielectric spheres with $\varepsilon_r = 2$ at different size-to-wavelength ratios $a/\lambda$.}
\begin{tabular}{cccccc}
\hline
\textbf{$a/\lambda$} & \textbf{$L/a$} & \textbf{$a/\Delta x$} & \textbf{$\lambda_{\varepsilon_r}/\Delta x$} & \textbf{Thread} & \textbf{Time (hrs)} \\
\hline
0.1 & 10    & 30    & 212    & 20   & 3.8   \\
1    & 4     & 71    & 50     & 10   & 2.8   \\
2    & 3     & 141   & 50     & 10   & 9.4   \\
\hline
\label{table:dielectric_sphere_er_2}
\end{tabular}
\end{table}

The computational domain sizes, grid resolutions, and execution times are summarized in Table~\ref{table:dielectric_sphere_er_2}. Despite the increased computational cost associated with fully three-dimensional simulations, the results demonstrate that the LBM can reproduce the main scattering characteristics of dielectric spheres over the range of size-to-wavelength ratios considered.

\section{Conclusion} \label{sec:conclusion}

In this work, we have demonstrated the applicability of the LBM to electromagnetic wave scattering from both canonical and non-canonical geometries. The method was validated against analytical and semi-analytical benchmarks over a range of size-to-wavelength ratios and dielectric constants. For two-dimensional configurations, including PEC and dielectric circular cylinders as well as dielectric hexagonal cylinders, results were presented for $a/\lambda$ ranging from $0.1$ to $10$.

High-permittivity cases for dielectric circular cylinders, with $\varepsilon_r$ up to $20$, were also examined. The results show that the LBM remains accurate for strong dielectric contrasts within the range considered. For hexagonal cylinders, comparisons with the DMF demonstrate that diffraction and edge effects associated with sharp geometrical features are well captured. For dielectric spheres, good agreement is obtained for small and moderate size-to-wavelength ratios, while larger values of $a/\lambda$ lead to increasing discrepancies due to resolution limitations in three-dimensional simulations. These discrepancies reflect the significantly higher spatial resolution required to resolve electromagnetic fields in three dimensions, resulting in increased computational cost.

From a methodological perspective, LBM provides a time-domain solution of Maxwell’s equations based on a kinetic formulation involving distribution functions, in contrast to conventional field-based discretizations such as FDTD. While both approaches employ explicit time stepping on structured grids, LBM offers an alternative numerical framework that has been shown to exhibit favorable properties in certain regimes, including reduced numerical dispersion and improved long-time stability, albeit at increased computational cost for a given resolution.

Accordingly, LBM should be viewed as a complementary approach rather than a replacement for established methods such as FDTD, FEM, or DDA. Its characteristics make it particularly attractive for transient wave problems and scenarios where numerical dispersion and long-time stability are critical.

The kinetic formulation of LBM also facilitates extensions to multiphysics problems and complex material models. In particular, its origin in fluid dynamics enables natural coupling with fluid flow, heat transfer, or particle transport within a unified computational framework. However, these capabilities remain to be explored systematically.

Future work will focus on improving accuracy for large size-to-wavelength ratios, especially in three-dimensional simulations, and on systematic comparisons with established numerical methods in terms of accuracy and computational efficiency. Further developments are also needed in boundary condition treatments, including total-field/scattered-field formulations and advanced absorbing layers analogous to perfectly matched layers in FDTD. In addition, incorporating material absorption will be essential for extending the method to more realistic scenarios. Finally, the reported computational costs depend on implementation details and should not be interpreted as definitive performance comparisons.

\section{Supplementary Material}

We are creating an open-source LBM solver for electromagnetic wave scattering and radiation force calculations, using the same code for all the analyses presented in this paper. You can access the code at the following link: \href{https://github.com/mohd-meraj-khan/LBM-for-scattering}{https://github.com/mohd-meraj-khan/LBM-for-scattering}.

\section*{Declaration of generative AI and AI-assisted technologies in the manuscript preparation process}

During the preparation of this work, the authors used \textit{ChatGPT} (OpenAI) to assist with language refinement, paraphrasing, and \LaTeX{} formatting tasks such as generating TikZ code for schematics. After using this tool, the authors reviewed and edited the content as needed and take full responsibility for the content of the published article.

\bibliographystyle{elsarticle-num} 
\bibliography{reference}

\end{document}